\documentclass[9pt,journal]{IEEEtran}

\usepackage[backend=bibtex,style=ieee,natbib=true]{biblatex}
\usepackage{hhline}
\usepackage{multirow}
\usepackage{mathtools}
\usepackage[table]{xcolor}
\usepackage{xtab}
\usepackage{afterpage}
\usepackage{graphicx}
\usepackage{tcolorbox}

\usepackage{relsize}
\usepackage{hyperref}
\usepackage{balance}
\usepackage{import}
\usepackage{cleveref}

\usepackage{booktabs}

\usepackage{caption}
\usepackage{fontawesome5}
\graphicspath{{./images}}

\usepackage{pdflscape}

\usepackage{enumitem} 

\usepackage{tikz}
\usepackage{varwidth}
\usepackage[framemethod=TikZ]{mdframed}

\usepackage[shortcuts]{extdash} 

\begin{document}

\usetikzlibrary{patterns}



\newenvironment{textboxtemplate}[3][]{%
  \par\nobreak\penalty0
  \vtop\bgroup%
  \ifstrempty{#1}
    {\mdfsetup{frametitle={}}}%
        {%
            \mdfsetup{innerleftmargin=10pt, innerrightmargin=10pt,
                frametitle={%
                    \tikz[baseline=(current bounding box.east),
                    outer sep=0pt]
                    \node[anchor=east,rectangle,draw=black!75,fill=#3]
                    {\textcolor{black}{\begin{varwidth}{\linewidth-10pt}~\centering#1\end{varwidth}~}};
                }%
            }%
    }%
  \mdfsetup{innertopmargin=4pt,innerbottommargin=5pt, linecolor=black!75,%
      linewidth=.5pt,topline=true,roundcorner=5pt,skipabove=5pt,
    frametitleaboveskip=\dimexpr-\ht\strutbox\relax}
  \begin{mdframed}[]\relax%
  \label{#2}%
  \setlength{\parindent}{0pt}
}
{%
  \end{mdframed}%
  \par\xdef\tpd{\the\prevdepth}\egroup
  \prevdepth=\tpd%
  }

\newenvironment{textbox}[2][]{%
    \begin{textboxtemplate}[#1]{#2}{tablebg}
  }{
    \end{textboxtemplate}
}

\newenvironment{interviewbox}[2]{
    \begin{textboxtemplate}[#1]{interviewexcerpt:#2}{tablebg}
    }{
\end{textboxtemplate}
}

\newcounter{hyp}
\newenvironment{hypothesisbox}{
    \refstepcounter{hyp}%
    \begin{textboxtemplate}[\faIcon{lightbulb} Hypothesis {\thehyp}]{hyp:\thehyp}{tablebg}
    }{
\end{textboxtemplate}}

\newenvironment{questionbox}[2]{
  \begin{textboxtemplate}[\faIcon{question-circle} RQ{#2}: {#1}]{answer:#2}{tablebg}
    }{
    \end{textboxtemplate}}


\definecolor{pastelgreen}{HTML}{ccebc5}
\definecolor{pastelyellow}{HTML}{fed9a6}
\definecolor{pastelred}{HTML}{fbb4ae}
\definecolor{pastelrose}{HTML}{fddaec}
\definecolor{pastelgray}{HTML}{f2f2f2}
\definecolor{tablebg}{HTML}{f0f0f0}

\definecolor{firefoxrose}{HTML}{de0f63}


\newcommand{\stackoverflow}{Stack Overflow}

\newcommand{\markmark}[1]{{\color{firefoxrose} Mark: #1}}
\newcommand{\markasks}[1]{{\color{firefoxrose} Question from Mark: #1}}
\newcommand{\andy}[1]{{\color{orange} Andy: #1}}
\newcommand{\rashina}[1]{{\color{red} Rashina: #1}}
\newcommand{\td}[1]{{\color{firefoxrose} #1}} 
\newcommand{\caro}[1]{{\color{blue} Caro: #1}}

\newcommand{\loabw}[2][to do: wording]{{\color{red} #2 (#1)}}

\newcommand*\circled[2][tablebg]{\tikz[baseline=(char.base)]{
\node[minimum size=\baselineskip,shape=circle,draw,inner sep=1pt,font=\footnotesize,fill=#1,] (char) {#2};}}

\newcommand\freefootnote[1]{%
  \begingroup
  \renewcommand\thefootnote{}\footnote{#1}%
  \addtocounter{footnote}{-1}%
  \endgroup
}

\newcounter{subj}
\refstepcounter{subj}
\makeatletter
\newcommand\subj[3]{%
    \phantomsection%
{\texttt{#3}}\def\@currentlabel{\unexpanded{#3}}\label{subj:#2}}%
\makeatother

\newcommand\interviewq[3]{%
    \begin{interviewbox}{\faIcon{comments} Interview~\ref{subj:#2} at {#3}}{}
        #1
    \end{interviewbox}}

\newcommand\code[1]{\textit{\##1}}

\newcommand\interviewer[1]{{\footnotesize\faIcon{microphone}}:~``\textit{#1}''}
\newcommand\intervieweenoref[1]{{\footnotesize\faIcon{comment}}~``\textit{#1}''}
\newcommand\interviewee[3]{\intervieweenoref{#1} (\ref{subj:#2} at #3)}

\newcounter{insight}
\newcommand\insight[1]{%
    \refstepcounter{insight}%
    \begin{textbox}[]{insight:\theinsight}
        #1
    \end{textbox}
}

\newcommand\subcategory[1]{\textbf{#1}}

\newcommand\weakevidence[0]{{\small\faIcon{user}}}
\newcommand\evidence[0]{\faIcon{user-friends}}
\newcommand\strongevidence[0]{\faIcon{users}}

\newcommand\weblink[2]{\href{#1}{\faIcon{link}~#2}}

\title{%
An instrument to measure factors that constitute \\
the socio-technical context of testing experience}


\author{\IEEEauthorblockN{Mark Swillus\IEEEauthorrefmark{1},
Carolin Brandt\IEEEauthorrefmark{2} and Andy Zaidman\IEEEauthorrefmark{3}} \\
\IEEEauthorblockA{
Delft University of Technology\\
The Netherlands\\
Email: \IEEEauthorrefmark{1}m.swillus@tudelft.nl,
\IEEEauthorrefmark{2}c.e.brandt@tudelft.nl,
\IEEEauthorrefmark{3}a.e.zaidman@tudelft.nl}}

\maketitle
\begin{abstract}
We consider testing a cooperative and social practice
that is shaped by the tools developers use,
the tests they write,
and their mindsets and human needs.
This work is one part of a project that
explores the human-
and socio-technical context
of testing
through the lens of those interwoven elements:
test suite and tools as technical infrastructure
and collaborative factors and motivation as mindset.
Drawing on empirical observations of previous work,
this survey examines how these factors relate to each other.
We want to understand which combination of
factors can help developers strive and make the most of their ambitions
to leverage the potential that software testing practices have.
In this report, we construct a survey instrument to measure
the factors that constitute the socio-technical
context of testing experience.
In addition,
we state our hypotheses about how these factors impact testing experience
and explain the 
considerations and process that led to the construction
of the survey questions.
\end{abstract}

\section{Introduction}
Software testing practices, which we define as the systematic
usage of software development tools to automate
the verification process of software,
is an effective way to prevent unexpected failures of software systems
which can have a detrimental effect on people and society.
For example, in 2024, a software bug that co-occurred
with a software testing system bug~\citep{pariseau_crowdstrike_2024},
grounded flights at airports around the world
and even disrupted hospital operation~\citep{pilkington_us_2024}.
Many researchers and practitioners
also recognize benefits that go beyond failure prevention~\cite{dudekula_mohammad_rafi_benefits_2012}.
Testing can facilitate software development processes and
is considered as a core component
in software development methodologies like extreme programming~\cite{wells_extremeprogrammingorg_2009}.
Given its impact on both software quality and the development process,
academics have been encouraging research on software testing practices for more than 40 years.
And research in this area continues to be exciting.
Recent advancements in the field of AI,
especially with the usage of large language models for code generation
have provoked new approaches to testing.

Today, there is a vast ecosystem of concepts, tools, and approaches
for software testing.
Given this diversity, one would think
every developer is able to integrate testing practices
into their projects in a way that suits their needs.
However, research into testing experience (TX)
paints a different picture.
Developers demonstrate emotional responses to testing~\citep{evans_scared_2021},
seem discouraged from meeting their own testing ambitions~\citep{swillus_sentiment_2023},
and even perceive it as a daunting task~\citep{stobie_too_2005}.
Despite the widely recognized benefits
of testing and the availability of tools and techniques
for almost every use case,
it is still seen as optional.
Especially when time is short and deadlines are near~\citep{wiklund_impediments_2017}.
As researchers based in the Netherlands,
we find the comparison to helmet use among Dutch cyclists fitting:
Studies have demonstrated that helmets save lives time and again~\citep{buth_effectiveness_2023}.
The advantages and importance of helmet use are clear,
and yet most Dutch cyclists,
confident in their habits and environment,
choose to ride without.
Further, in bordering countries like Germany,
where cultural and infrastructural factors differ,
helmet usage is more common.
The reasons for not wearing helmets in the Netherlands,
just like the choice of developers not to test, we argue,
are not of a purely objective
or technical nature.
Both phenomena can only be understood by considering the broader context in which they occur.
In the case of cycling, this broader context concerns practical aspects
like infrastructure or road safety, and personal concerns like the leisureliness,
confidence, and joy with which people use their bicycles.
Our prior work explores the broad context that influences testing decisions,
and we find that in the case of testing,
practical aspects and personal concerns play a role as well~\cite{swillus_who_2025}.
A combination of various technical and non-technical factors,
such as the availability of testing infrastructure (tools and test automation)
and the presence of a \textit{testing culture} seem to play an important role.
Adoption and adaptation of testing strategies
does not seem to be a linear process.
Rather, it can better be described as a collective pushing of ambitions,
ideas, and technical possibilities through a
filter of available materials and techniques.
In this report, we continue our investigation of this entanglement.
We first propose hypotheses about the
conditions under which testing is enabled or inhibited,
and then present a survey instrument with which
we are going to test those hypotheses.
By publishing our research design and survey instrument ahead of data collection,
we demonstrate transparency and methodological rigor.
Publishing preconceptions and intentions in advance
precludes practices
like HARKing (hypothesizing after results are known),
which can compromise the integrity of research projects and the validity of findings.

\subsection{Pre-registration of hypotheses and research design}
Studies like the recent work of~\citet{cologna_trust_2025},
who investigate and discuss public trust in scientists,
echo what can be observed in contemporary public discourse
and policymaking:
if a fraction of society loses its trust in scientific work,
the consequences, like in the case of COVID-19, 
can be disastrous~\cite{hamilton_elite_2021}.
According to~\citet{cologna_trust_2025},
the majority of people still consider science
an important foundation for change.
However,
rapid developments in artificial intelligence
have raised new concerns about the
integrity of research and the reliability of research findings~\cite{Elali_AI_generated_2023}.
In the context of these developments,
we are motivated
to demonstrate transparency, rigor, and integrity in our work.
Engaging in practices promoted by open science initiatives
is one way to act on that motivation.
By making the research process and results more accessible and transparent,
open science initiatives improve reproducibility and foster greater trust in scientific work.
In the discipline of software engineering, this is often done by publishing replication packages, which include
datasets and source code when the results of a project are published.
However, transparency can begin earlier
with the publication of research design and instruments,
before data is collected and analyzed.
For example, survey instruments can be shared prior to data collection
to allow other researchers to assess or reuse them in different contexts,
independent of the results of the research project they were constructed for (see~\cite{dyba_instrument_2000}).
In line with this practice, we are publishing the present report,
which includes the survey instrument we intend to use before collecting and analyzing data.

In empirical software engineering research, the publication of research designs is becoming more common.
Some conferences and journals now offer registered reports tracks~\cite{ICSE_msr_2025, EMSE_registered_2025},
where the study design undergoes a peer-review process before data is gathered.
Early feedback and transparency help improve methodological rigor, and in many cases,
this feedback also comes with a commitment to publish the study regardless of its results.
On the one hand, this discourages practices like HARKing (hypothesizing after results are known),
which can compromise the validity of findings.
On the other, it encourages the publication of negative results,
which are often undervalued but critical to advancing knowledge in the field.
Not being pressured to produce positive outcomes to secure publication,
researchers are more likely to uphold the scientific imperative of disinterestedness.

By publishing our research design ahead of data collection,
we act on our motivation to strive for a more open and trustworthy research culture
in empirical software engineering.
This report can therefore also be understood as a commitment to
publish outcomes independent of their implications for our prior work.

\section{Method}
Our prior exploratory work investigated
testing experiences using qualitative research strategies.
In~\cite{swillus_sentiment_2023},
we analyze documents to uncover the roots of sentimental attitudes of developers towards testing.
Following that, in~\citep{swillus_who_2025},
we analyzed semi-structured interviews to explore the topic further.
Data triangulation enables researchers to understand a phenomenon from multiple perspectives.
Method triangulation reduces the bias
that using a single method
to compare different perspectives
can introduce~\cite{storey_who_2020}.
We therefore choose a survey-based study design
to deepen our work on socio-technical aspects of software testing.
In the following sections, we describe how we construct
the survey instrument.
We follow~\citet{kitchenham_personal_2008}, which provides
a guideline to create personal opinion surveys.
The rest of this section is organized according to the six
activities Kitchenham et al.\ identify in their guideline.
As this report is published prior to data collection,
we only consider the first four activities, which lead to
the construction of a survey instrument
and a preliminary plan for its evaluation.

\begin{itemize}
\item Setting the objective
\item Survey Design
\item Developing the survey instrument
\item Evaluating the survey instrument
\end{itemize}

\section{Setting the Objective}

Declaring the research objective, including the hoped-for outcomes, is necessary to constrain the scope of questions in a survey~\cite{kitchenham_principles_2002}.
To identify the research objective,
we choose a top-down approach~\cite[p.10]{Linaker_guidelines_2015},
breaking down the broader research question we want to tackle
into concrete hypotheses based on prior work.

\subsection{Research Questions}
We argue that the many factors influencing developers'
engagement with testing often lead to unanticipated outcomes,
which can understandably provoke emotions.
We align with~\citet{rooksby_testing_2009} who describe testing as a stochastic process:
What makes a test plan effective is often the ability to handle various
unpredictable contingencies as they arise.
In our prior work~\cite{swillus_who_2025},
we identify factors that can give rise
to these contingencies,
and others that
enable developers to manage them.
The objective of this study is to further
investigate the interplay of those factors:

\begin{textbox}{}
\begin{itemize}
  \item[\textbf{\texttt{RQ1}}] How does the interplay of testing conditions shape a software developer's motivation to use testing?
  \item[\textbf{\texttt{RQ2}}] How does the interplay of testing conditions shape a software developer's perception of the value of testing?
\end{itemize}
\end{textbox}

Through our preliminary work,
we observed that concrete conditions influence how testing is perceived and used,
as well as which factors encourage
developers to reflect on their testing practices.
While we have identified conditions related to developers' motivation to test, we have yet to determine
whether there are common patterns or combinations
that stimulate or inhibit testing.
A key goal of our work is to identify these relationships and examine how they correlate with testing usage and motivation.

\begin{textbox}{}
\begin{itemize}
  \item[\textbf{\texttt{RQ3}}] Which sets of conditions are common among developers who are motivated to engage in testing?
\end{itemize}
\end{textbox}

To investigate this question,
we aim to measure developers' motivation regarding software testing,
their self-reported perception of effort spent on testing, and
their evaluation of how conditions such as \textit{testing infrastructure},
\textit{complexity}, or a developer's \textit{autonomy}
impact their efforts.
We then test a set of hypotheses
that emerged from our previous qualitative exploratory work.

\section{Survey Design}

We choose to construct a descriptive cross-sectional survey
which aims to discover factors
that affect software testing (predispositions) and their relationship with each other.
Participants are asked for information at one fixed point in time.
The survey therefore provides a snapshot of what developers experience.
We implement the survey using a web-based self-administered questionnaire,
as this format serves both our need for method triangulation and facilitates
simple distribution and coverage of the population we want to study.

Before we approach the implementation of the survey instrument,
we conduct a lean literature review to determine how others have investigated
the factors we want to measure with our instrument.
Re-using tested and peer-reviewed instruments increases the credibility and reproducibility
of results.

\section{Lean literature review}\label{sec:background}

By reviewing available literature,
we establish to what extent
the construction of a new instrument is required.
Re-using tested and peer-reviewed instruments
increases the credibility and reproducibility of results.
We also familiarize ourselves with approaches others have chosen
to investigate similar phenomena,
in order to learn how the many aspects of it can be observed,
conceptualized, and measured.
Identifying measurable variables through this process
is the first step in constructing
survey questions.
The aim of a lean literature review is not to provide a comprehensive overview
of all relevant topics, but rather to scan and scope,
optimized to serve its key aims quickly and as easily as possible~\cite[p.119]{hoda_qualitative_2024}.

\subsubsection{Motivation}

Motivation of software developers has already been researched for over 40 years,
as it has been identified as an important factor impacting software developer productivity.
Questionnaires are often used to measure various aspects that relate to motivation~\cite{beecham_motivation_2008}.
Over the years, many different motivators that affect developers, including
recognition, trust, autonomy, and variety of work, have been identified~\cite{beecham_motivation_2008, franca_motivation_2011}.
Different models of motivation of software developers have been proposed~\cite{sharp_models_2009},
and though many different perspectives on motivation have been researched,
it remains a complex topic.
As the world is changing, past work on motivation needs re-validation to remain insightful.
Unfortunately this is done very little~\cite{sharp_models_2009}.
A more recent study
by \citet{verner_factors_2014}
about motivation using survey instruments
finds that social factors and human needs are important for developer motivation.
Developers seem to be motivated
by a project manager with good communication skills,
in projects where risk is controlled,
when the work environment is supportive,
In their survey,
\citet{verner_factors_2014}
measure motivation by asking developers
in a self-administered online questionnaire
about the motivation of their team members.
Similar to Daka and Fraser 
who asking participants directly:
\textit{"What motivates you to write unit tests?"}~\cite{daka_survey_2014}.
Straubinger et al., who investigate opinions of software developers regarding testing,
take a different approach
because they identify that \textit{motivation} is an overloaded term~\cite{straubinger_survey_2023}.
Motivation, they argue, is a multi-dimensional concept
which needs to be considered from many angles
in order to be analysed.
They do not directly ask developers about their motivation but
instead ask about circumstances of their work that transcend motivation.
Aligning with this view,
social psychologists consider motivation to be a construct that
cannot be recorded or observed directly.
Accordingly, measuring motivation in experiments or through questionnaires
is considered a non-trivial task.
In a guideline to measure motivation,
Touré‐Tillery et al. suggest differentiating between
outcome-focused motivation to complete a goal,
and process-focused motivation
(more commonly known as intrinsic motivation),
which has less emphasis on the outcome and includes elements
such as appropriate means and enjoyment during
goal pursuit~\cite{touretillery_how_2014}.
The first resembles an attitude of \textit{getting it done}, the latter of \textit{doing it right}.
In experiments, indicators for those two aspects
can be measured using different means.
For example, outcome-focused motivation can be revealed through
a subject's positive evaluation of goal-congruent constructs
such as means, objects, or persons, and a subject's negative evaluation
of goal-unrelated constructs such as distractions.
Process-focused motivation is, in this context, revealed by positive
evaluation of the process~\cite{touretillery_how_2014}.

\subsubsection{Human factors and software development processes}

Several recent studies investigate how human and social
factors influence software development practices
and the organization of software development projects.
Investigating success factors of agile method adoption
using a survey instrument, Misra et al. find that
technical competency, team size, and planning are not strong factors,
but that corporate culture and training initiatives have a significant effect~\cite{misra_identifying_2009}.
Instead of measuring success factors like \textit{Corporate culture} using
with questions, each success factor was measured using a set of questions
that collectively represent the factor.
Dybå uses a similar approach in a questionnaire-based
study investigating the impact of company size on software process improvement~\cite{dyba_factors_2003}.
According to their study, the size of companies in which software
is developed influences how well projects can leverage different kinds of knowledge and expertise.
To investigate influences on software process improvement,
Dybå identifies six key success factors
and breaks each factor down into several indicators, each measured
using one question.
The approach is motivated by the argument
that complex concepts like software process improvement success factors
can not be reliably measured through simple one-dimensional questions.
Instead of measuring a factor with one question,
multiple questions are therefore
combined to measure each concept~\cite{dyba_factors_2003}.
A similar approach is taken by Machuca-Villegas et al.,
who investigate the perception of human factors
that influence the productivity of software development teams~\cite{machuca_villegas_perceptions_2022}.
In their survey instrument, they
first consider key human factors
and then scrutinize each using several questions~\citep{machuca_villegas_instrument_2021}.
Using the perspective of human factors, they
find a positive influence of empathy, social interaction, communication, and autonomy on productivity.

\subsubsection{Software testing: Opinions and needs of developers}

In an attempt to close a gap between academic and practitioner views on
software testing, Rafi et al. use a combination of a systematic literature review
and a practitioner survey to investigate benefits
and limitations of software testing practices~\citep{dudekula_mohammad_rafi_benefits_2012}.
Using their literature review, they derive a set of hypotheses
which they test by asking practitioners whether they agree or disagree.
As the studies considered in their review mostly focus on technical aspects,
the results of their study also primarily concern technical aspects of software testing.
For example, their survey demonstrates that the benefits of test
automation are related to test reusability,
repeatability, test coverage, and effort saved in test execution~\citep{dudekula_mohammad_rafi_benefits_2012}.
\citet{daka_survey_2014} also investigate testing practices with a focus on established
practices and problems.
They identify that there is a need to improve the technical capabilities
of unit testing, especially in terms of automation.
However, they also find that the need for testing is often motivated
organizationally, and that a developer's own conviction is a strong factor.
Like Rafi et al.~\citep{dudekula_mohammad_rafi_benefits_2012},
\citet{daka_survey_2014} measure aspects with distinct questions,
but to increase the validity of those questions,
they use verification questions—questions that 
measure the same variable using alternate wording.
They then disregard responses where answers to the same question diverge.
All questions were derived from their research questions,
but how exactly those questions were developed is not clearly explained~\cite{daka_survey_2014}.

\section{Hypotheses and Variables}\label{sec:hypotheses}

In our prior work~\cite{swillus_who_2025},
we identify eleven conditions
that affect developers' testing practices:

\begin{itemize}
\item Complexity
\item Software Development Process
\item Safety \& Responsibility
\item Business and Application Domain
\item Vision
\item Resource Usage
\item Mandates
\item Testing Infrastructure
\item Testing Culture
\item Community Perspective
\item Personal Leanings
\end{itemize}

We also proposed a theory that conceptualizes software testing as a dualism, composed of ephemeral and material
elements that influence each other \cite[see][pp.17-22]{swillus_who_2025}.
To approach the research questions we above,
we focus our investigation on the interplay of a subset of above-mentioned conditions.
First, we investigate how organizational conditions relate to
developers' motivation to use testing practices and the extent to which testing is used in projects.
Second, we explore the relationship between complexity and testing infrastructure in relation to testing motivation.
Third, we examine how material and ephemeral aspects of testing influence one another.

\subsubsection{Software Process and Testing Motivation}

The introduction or adaptation of software testing methods
within an organization can be seen as part of an effort to
improve the software development process.
\citet{dyba_factors_2003} identified that
the success of process improvement efforts depends on the correlation
of particular organizational factors, including company size.
Large companies benefit from structured processes that leverage past knowledge,
while smaller organizations benefit more from experimentation and exploration~\cite{dyba_factors_2003}.
Our prior research suggests that business context and
software development organization
influence testing practices as well~\cite{swillus_who_2025, rooksby_testing_2009}.
Autonomy (reflected in the capacity to experiment)
and the exploitation of existing knowledge
appear to play major roles.
Based on these findings,
we propose the following hypotheses about the relationship between autonomy,
knowledge sharing, and company size in the context of testing:

\begin{textbox}{}
\begin{itemize}
  \item[\textbf{\texttt{[H1]}}] Organizational factors affect testing practices
\begin{itemize}
  \item[\textbf{\texttt{H1}.\texttt{1}}] In large companies, testing is used more extensively when past knowledge is extensively leveraged
  \item[\textbf{\texttt{H1}.\texttt{2}}] In small companies, testing is used more extensively when new ideas and exploratory approaches are embraced
\end{itemize}
\item[\textbf{\texttt{[H2]}}] Motivation to test (both goal-focused and process-focused) stems mainly from individual conviction and not from organizational factors. However, organizational factors do impact the extent to which testing is used.
\begin{itemize}
\item[\textbf{\texttt{H2}.\texttt{1}}] Motivation to test is not significantly influenced by organizational factors
  \item[\textbf{\texttt{H2}.\texttt{2}}] The extent of testing is influenced by organizational factors such as business sector, mandates and company size
\end{itemize}
\end{itemize}

\end{textbox}

Testing these hypotheses
requires measuring the following variables,
for some of which measurement methods are described
in the literature we reviewed in \Cref{sec:background}:

\begin{itemize}
  \item Company size (\texttt{H1})
  \item Business sector (\texttt{H2})
  \item Employee participation~\cite{dyba_instrument_2000} and Mandates (\texttt{H2})
  \item Extent of testing (\texttt{H1})
  \item Exploitation of existing knowledge~\cite{dyba_instrument_2000} (\texttt{H2})
  \item Exploration of new knowledge~\cite{dyba_instrument_2000} (\texttt{H2})
  \item Motivation (process-focused and goal-focused)~\cite{touretillery_how_2014} (\texttt{H2})
\end{itemize}

\subsubsection{Complexity and Testing Infrastructure}

The complexity of a software system can influence testing motivation in several ways.
When software consists of multiple interacting components,
systematic testing can help developers maintain an overview of the project.
Testing becomes a way to reduce perceived complexity.
However, complexity can also discourage testing if simple approaches
are inadequate and the effort required for effective testing becomes overwhelming~\cite{swillus_sentiment_2023}.
We argue that the availability of testing infrastructure
and team culture significantly mediate this relationship:

\begin{textbox}{}
\begin{itemize}
  \item[\textbf{\texttt{[H3]}}] The presence of usable testing infrastructure increases the process-oriented motivation to test
  \item[\textbf{\texttt{[H4]}}] The (perceived) complexity of a project impacts motivation to test
\begin{itemize}
  \item[\textbf{\texttt{H4}.\texttt{1}}] Complexity decreases process-oriented testing motivation
  \item[\textbf{\texttt{H4}.\texttt{2}}] Complexity increases process-oriented motivation and goal-oriented motivation if testing infrastructure is present 
  \item[\textbf{\texttt{H4}.\texttt{3}}] Complexity decreases process-focused motivation if testing infrastructure is absent
\end{itemize}
\end{itemize}
\end{textbox}

Testing these hypotheses
requires measuring the following variables,
for some of which measurement methods are described
in the literature we reviewed in \Cref{sec:background}:

\begin{itemize}
  \item Testing infrastructure (\texttt{H3}, \texttt{H4})
  \item Motivation (process-focused and goal-focused)~\cite{touretillery_how_2014} (\texttt{H3}, \texttt{H4})
  \item Complexity (\texttt{H4})
\end{itemize}

\subsubsection{Material and Social Construction of Testing}

In prior work\cite{swillus_who_2025},
we conceptualized software testing
as an interplay between material elements
(e.g., test frameworks, source code, documentation)
and ephemeral elements
(e.g., discussions, culture).
Developers leave \textit{signatures} in code
that reflect their testing values,
while discussions and team culture produce
\textit{echoes} that foster testing culture.
Testing-signatures can be understood
as traces in artifacts (e.g., testing infrastructure) that represent
developers' knowledge and experience and can spark reflection
when encountered.
We argue that the presence of this
record of experience and knowledge can,
when properly exploited by an organization
stimulate discussions and reflections on testing.

Testing-echoes are collaborative reflections,
of ideas and knowledge in
the context of testing.
We argue that this collaborative exploration
of new ideas and a positive testing culture
leads to the extension
of testing efforts.
In this context, we argue that it is the process
of reflection that motivates developers to extend their testing efforts.

We further hypothesize that
ambitions to extend testing practices
benefit from autonomy,
meaning the developer has influence
on how the development process is structured.

\begin{textbox}{}

\begin{itemize}
  \item[\textbf{\texttt{[H5]}}] Knowledge exploration positively influences the extent of testing efforts
\begin{itemize}
  \item[\textbf{\texttt{H5}.\texttt{1}}] Active discussion and interaction influences the extent to which knowledge exploration impacts testing effort extension
\end{itemize}
\item[\textbf{\texttt{[H6]}}] Knowledge exploitation positively influences the extent to which testing practices are reflected
\begin{itemize}
  \item[\textbf{\texttt{H6}.\texttt{1}}] The presence of testing infrastructure influences the extent of reflection on testing
\end{itemize}
\item[\textbf{\texttt{[H7]}}] When testing practices are discussed interactively, motivation (goal-focused) is higher and adoption of testing practices is greater
\item[\textbf{\texttt{[H8]}}] Whether reflection through testing echoes leads to code changes depends on the participant's level of autonomy
\end{itemize}
\end{textbox}

Testing these hypotheses
requires measuring the following variables,
for some of which measurement methods are described
in the literature we reviewed in \Cref{sec:background}:

\begin{itemize}
  \item Exploration of new knowledge \cite{dyba_instrument_2000} (\texttt{H5})
  \item Extent of testing (\texttt{H5}, \texttt{H7})
  \item Discussion and interaction (\texttt{H5}, \texttt{H6}, \texttt{H8})
  \item Exploitation of existing knowledge \cite{dyba_instrument_2000} (\texttt{H6})
  \item Testing infrastructure (\texttt{H6})
  \item Motivation (goal-focused)~\cite{touretillery_how_2014} (\texttt{H7})
  \item Employee participation~\cite{dyba_instrument_2000} and Mandates (\texttt{H8})
\end{itemize}

\subsection{Summary}

We aim to illuminate relationships
between organizational, technical, social, and cultural
factors that shape software testing practices.
The survey instrument we construct to
investigate these relationships is designed to
deepen our understanding
of how and why developers
are motivated to adopt testing,
while also providing actionable
insights to enhance testing
processes across diverse development environments.
In order to reach our research objective we measure
10 variables:

\begin{enumerate}
  \item Company size (\texttt{H1})
  \item Business sector (\texttt{H2})
  \item Extent of testing (\texttt{H1}, \texttt{H5}, \texttt{H7})
  \item Complexity (\texttt{H4})
  \item Testing infrastructure (\texttt{H3}, \texttt{H4}, \texttt{H6})
  \item Discussion and interaction (\texttt{H5}, \texttt{H6}, \texttt{H8})
  \item Employee participation~\cite{dyba_instrument_2000} and Mandates (\texttt{H2}, \texttt{H8})
  \item Exploitation of existing knowledge~\cite{dyba_instrument_2000} (\texttt{H2}, \texttt{H6})
  \item Exploration of new knowledge~\cite{dyba_instrument_2000} (\texttt{H2}, \texttt{H5})
  \item Motivation (process-focused and goal-focused)~\cite{touretillery_how_2014} (\texttt{H2}, \texttt{H3}, \texttt{H4}, \texttt{H7})
\end{enumerate}

Using a questionnaire to measure those variables 
we test eight hypotheses and aim to generate
insights that enable
more effective integration
of testing strategies,
ensuring that tools and
techniques align with developers'
needs, goals, and the broader
organizational contexts in which they are employed.

%
\section{Implementation}

For the 10 variables we want to measure with a questionnaire,
we re-use three key factors of software process improvement in organizations
from \citet{dyba_instrument_2000}.

We adapt the questions were needed to fit our focus on software testing.
To measure motivation, we use indicators
for process-focused and goal-focused motivation as suggested by~\cite{touretillery_how_2014}.
To measure goal-focused motivation we construct 5 questions which
address the recall and evaluation of goal-related constructs that relate to testing.
To measure process-focused motivation we construct 3 questions which
address the evaluation of the impact of testing on the process of software development.

Of the other six questions,
we measure two variables (company size and business sector)
with single-item questions as they are similar to demographical questions
and therefore unambiguous.
For company size, a range
from a set of fixed values can be selected
(e.g. 2-10, 11-50, 51-200, 201-500, 501+).
Similarly, we use the list from the Global Industry Classification Standard (GICS\footnote{\weblink{https://www.msci.com/our-solutions/indexes/gics}{msci.com/our-solutions/indexes/gics}}) 
from which participants can select the Industry sector that best fits their organization.

Both our prior work and the literature review presented above
indicate that the remaining variables we want to measure
are inherently complex and multi-faceted.
To reliably measure complex variables,
we therefore follow an approach similar to~\cite{dyba_instrument_2000}.
Instead of using a single question to measure the four remaining,
complex variables we use
multi-item scales which measure a single variable
with multiple representative indicators.
The scores (in this context called \textit{scale scores})
of each variable are then calculated through the
sum of all respective indicators.
For each factor we want to measure, we therefore construct
three or more indicators.
For each indicator we then construct one question.

We derive indicators for each variable
with our prior work on software testing experience~\cite{swillus_who_2025}.
We then construct questions by
turning each indicator into one unambiguous statement.
For example, to measure \textit{complexity},
one indicator is a developers'
\textit{perception of technical complexity of the project}.
From this indicator, we derive the following statement,
which is answered using a 5-point Likert scale
ranging from strongly agree to strongly disagree:
\textit{I consider the software project as complex: it is large and consists of many components which interact with each other}.
We choose 5-point Likert scales with the same
options for all questions in our survey instrument,
except for demographic questions.
\Cref{tab:questions} contains all factors and corresponding questions.

\tablecaption{Multi-scale variables and corresponding questions}
\tablefirsthead{%
\textbf{Variable}&\textbf{Indicator}&\textbf{Question} \\ \midrule}
\tablehead{%
\multicolumn{3}{c}%
{{Table continued from previous page}} \\[10pt]
\textbf{Variable}&\textbf{Indicator}&\textbf{Question} \\ \midrule}
\tabletail{%
\midrule \multicolumn{3}{r}{{Table continued on next page}} \\ \midrule}
\tablelasttail{%
\\\midrule
\multicolumn{3}{r}{{End of table}} \\ \bottomrule}

\xentrystretch{0.12} 
\afterpage{\onecolumn
\begin{mpxtabular}{lll}\label{tab:questions}
\multirow{8}{*}{Extent of Testing}      & Extent to which testing is used as an individual & I am making extensive use of software testing \\
                                        & activity to develop software                     & techniques in my daily software development activities \\
                                        & Extent to which testing is used by the project   & The project for which I develop software is making \\
                                        &                       to develop software        & extensive use of software testing \\
                                        & The organization supports and embraces software  & The organization in which the project is embedded                                                    \\
                                        & testing                                          & promotes the usage of software testing practices                                                     \\
                                        & Degree to which testing is mandated            & Testing is strictly required when contributing                                                       \\
                                        &                                                  & to the software I develop                                                                            \\ \hline
\multirow{6}{*}{Complexity}             & Perception of technical complexity of project    &   The software project I contribute to is large and                                                                                                     \\
                                        &                                                  & consists of many components which interact with each other                                           \\
                                        & Usage of standard tools                          & We mainly use common, standardized technologies when                                                 \\
                                        &                                                  & we develop, test, build and distribute the software we develop                                       \\ 
                                        & Degree to which best practices can be used       & Large parts of the code base use\\
                                        &                                                  & common patters that can be learned in books or online.\\ \hline
\multirow{8}{*}{Infrastructure}         & Extent of tool support for testing goals         & Our project is built upon strong testing frameworks                                                                                                     \\
                                        &                                                  & which include elements like test suites and development tools\\
                                        & Ability to use testing with respect to           & I (would) have enough time to expand and improve our\\
                                        & resources available                              & project's testing framework \\
                                        & Importance of existing source code for the       & I often reuse existing source code when developing\\
                                        & extensions of test suites                        & new tests\\
                                        & Extent of explicit guidelines for testing        & Guidelines and expectations for testing are clearly defined\\
                                        & in the project                                   & and explicitly documented \\ \hline
                                        & Degree to which discussions translate to         & Conversations about testing usually lead to actual\\
                                        & contributions                                    & improvements in our testing efforts \\
                                        & Degree to which testing is learned through       & I learned a lot about testing through colleagues who taught \\
                                        & mentoring                                        & me how to approach it \\
 \multirow{1}{*}{Discussion and}        & Extent of implicit guidelines for testing        & I mainly know what is expected of me in terms of testing \\
\multirow{1}{*}{Interaction}            & in the project                                   & by interacting and talking with other developers \\
                                        & Extent to which testing culture can establish    & There is a shared understanding in our project that parts \\
                                        & something as untestable                          & of the software we develop are untestable \\
                                        & Extent to which testing culture influences       & In our team testing tasks are considered to be \\
                                        & the perceived difficulty of testing in project   & straight forward and easily done \\ \hline
                                        & Extent of employee involvement in decisions      & Software developers are involved to a great extent \\
                                        & that should best be done at their own level      & in decisions about the implementation of their work \\
                                        & Extent to which employees contribute with        & Software developers are actively contributing with \\
Employee                                & improvement proposals                            & proposals to adapt testing strategies           \\
       Participation                    & Extent of developer participation in the         & Software developers are actively involved        \\
                                        & formalization of routines                        & in creating routines and procedures for testing  \\
                                        & Extent of ongoing dialog and discussion about    & We have an on-going dialogue and discussion about \\
                                        & software development                             & software development                             \\
                                        & Extent of ongoing dialog and discussion about    & We have an on-going dialogue and discussion about \\
                                        & software testing                                 & software testing \\ 
                                        & Extent to which existing knowledge is exploited  & We exploit the existing organizational knowledge to the utmost extent\\
Exploitation of                         & Extent of learning from past experiences         & We are systematically learning from the experience with prior projects\\
Existing                                & Degree to which formal routines are based on     & Our routines for software testing are based on   \\
Knowledge                               & past experience                                  & experience from prior projects \\
                                        & Degree of systematization of past experience     & We collect and classify experience from prior projects \\
                                        & Degree of internal experience transfer           & We put great emphasis on internal transfer of       \\
                                        &                                                  & positive and negative experience               \\ \hline 
                                        & Degree of adaptability to rapid change, increasing & We are very capable at managing uncertainty in the  \\ 
                                        & complexity and environmental uncertainty         & organization's environment                     \\ 
                                        & Extent to which innovation/change is encouraged  & In our organization, we encourage innovation and creativity \\
                                        & Degree of experimentation with new ideas,        & We often carry out trials with new software      \\
   Exploration of                       & strategies and technologies                      & engineering methods and tools \\
        New Knowledge                   & Degree of experimentation with new ideas,        & We often conduct experiments with new ways of    \\
                                        & strategies and technologies                      & working with software development methods. \\
                                        & Ability to question underlying values            & We have the ability to questions \textit{established} truths \\
                                        & Degree of flexibility in task execution          & We are very flexible in the way we carry our our work \\
                                        & Degree of detail in task specification           & We do not specify work processes more than what\\
                                        &                                                  & is absolutely necessary \\
                                        & Importance of matching the variety and           & We make the most of the diversity and interests to manage \\
                                        & complexity of the organization's environment     & the variety and complexity of the organizations environment \\ \hline
                                        & Extent to which testing is perceived to          & Software testing practices enable developers   \\
                                        & contribute to software quality goals             & to write better code    \\
                                        & Degree of accessibility and memory for           & I can name several testing tools and methods that \\
          Motivation                    & goal-congruent constructs (means, objects, person) & (would) help me to achieve the project's goals \\
          (goal-focused)                & Degree of accessibility and memory for           & I can recall many situations in which testing related \\
                                        & goal-incongruent constructs (temptations)        & work distracts me from getting the job done \\
                                        & Degree of positive evaluation of goal-congruent  & Testing tools, methods and testing contributions of my \\
                                        & constructs (means, objects, persons)             & colleagues are crucial for the success of our project \\
                                        & Degree of negative evaluation of goal-incongruent& Unnecessary testing tasks often distract me and hinder  \\
                                        & constructions (temptations, distractions)        & my progress towards my actual goals \\ \hline
                                        & Extent to which software testing contributes to pr. & Software testing increases my productivity     \\
       Motivation                       & Positive experience from the process             & I enjoy using software development more when I use \\
       (process-focused)                &                                                  & software testing tools and practices           \\
                                        & Positive evaluation of the process               & The use of software testing methods and tools \\
                                        &                                                  & positively impacts my workflow                 \\
\end{mpxtabular}
\twocolumn} 

\newpage
\section{Evaluating the Survey Instrument}
Before turning the instrument into a web-based questionnaire that can be
self-administered by participants,
all questions and answers should be evaluated.
Evaluation servers multiple goals\cite{kitchenham_personal_2008}

\begin{itemize}
  \item Ensure that questions are understandable
  \item Assess the likely response rate
  \item Determine the reliability and validity of the instrument
  \item Test the data analysis techniques on expected outcomes
\end{itemize}

To evaluate the survey instrument we first conduct an unstructured interview
with a software developer who we consider to be an expert on the topic.
After the interview in which we discuss the research objective
we incorporate the interviewee's feedback
into the first draft of our questionnaire
and then give our questionnaire the interviewee for a more structured evaluation.
We ask them to rate each question on a scale from one to three,
ranging from \textit{Not relevant - should be removed} to
\textit{relevant - can remain as is}
and provide explanations for each answer in free text fields.
We provide an extensive description of each point on the evaluation scale,
which we take from the work of Machuca-Villegas et al.~\citep{machuca_villegas_instrument_2021}
can be seen in \Cref{tab:evaluation}.

\begin{table}[]
\begin{tabular}{ll}
\textbf{Score}     & \textbf{Answer Criteria}                                                                                                                                  \\ \hline
\multirow{4}{*}{1} & • Not relevant so it should be removed                                                                                                                      \\
                   & • This item is not clear                                                                                                                                    \\
                   & • This item has no logical relationship with the factor. \\ 
                   & \begin{tabular}[c]{@{}l@{}}• This item can be removed without affecting the factor\\ measurement\end{tabular}                                               \\ \hline
\multirow{5}{*}{2} & • It must be rewritten                                                                                                                                      \\
                   & \begin{tabular}[c]{@{}l@{}}• The item requires several modifications or\\ a very large modification in terms of wording or structure\end{tabular}           \\
                   & • A very specific modification is required for some wording                                                                                                 \\
                   & \begin{tabular}[c]{@{}l@{}}• The item has a moderate relationship with the factor\\ it is measuring\end{tabular}                                            \\
                   & • The item is relatively important                                                                                                                          \\ \hline
\multirow{4}{*}{3} & • This item is relevant                                                                                                                                     \\
                   & • This item is clear with proper semantics and syntax                                                                                                       \\
                   & • This item is completely related to the factor being measured                                                                                              \\
                   & • This item is very relevant and must be included                                                                                                          
\end{tabular}
\caption{Item assessment criteria for survey instrument evaluation}
\label{tab:evaluation}
\end{table}

After we incorporate this in-depth feedback,
we conduct a pilot study for which we recruit a small group of
software developers to complete the questionnaire
and encourage them to provide feedback about questions using free-text fields.
Through feedback from the pilot study
we further improve wording and clarity where possible.
We also use the data gathered in the pilot study
to run preliminary tests for
reliability and validity.




\printbibliography 

@article{beecham_motivation_2008,
	title = {Motivation in {Software} {Engineering}: {A} systematic literature review},
	volume = {50},
	issn = {09505849},
	shorttitle = {Motivation in {Software} {Engineering}},
	url = {https://linkinghub.elsevier.com/retrieve/pii/S0950584907001097},
	doi = {10.1016/j.infsof.2007.09.004},
	language = {en},
	number = {9-10},
	urldate = {2021-06-24},
	journal = {Information and Software Technology},
	author = {Beecham, Sarah and Baddoo, Nathan and Hall, Tracy and Robinson, Hugh and Sharp, Helen},
	month = aug,
	year = {2008},
	pages = {860--878},
}

@article{wiklund_impediments_2017,
	title = {Impediments for software test automation: {A} systematic literature review: {Impediments} for {Software} {Test} {Automation}},
	volume = {27},
	issn = {09600833},
	shorttitle = {Impediments for software test automation},
	url = {https://onlinelibrary.wiley.com/doi/10.1002/stvr.1639},
	doi = {10.1002/stvr.1639},
	language = {en},
	number = {8},
	urldate = {2021-12-01},
	journal = {Software Testing, Verification and Reliability},
	author = {Wiklund, Kristian and Eldh, Sigrid and Sundmark, Daniel and Lundqvist, Kristina},
	month = dec,
	year = {2017},
	pages = {e1639},
}

@inproceedings{dudekula_mohammad_rafi_benefits_2012,
	title = {Benefits and limitations of automated software testing: {Systematic} literature review and practitioner survey},
	isbn = {978-1-4673-1822-8 978-1-4673-1821-1},
	shorttitle = {Benefits and limitations of automated software testing},
	url = {http://ieeexplore.ieee.org/document/6228988/},
	doi = {10.1109/IWAST.2012.6228988},
	urldate = {2021-12-01},
	booktitle = {2012 7th {International} {Workshop} on {Automation} of {Software} {Test} ({AST})},
	publisher = {IEEE},
	author = {{Dudekula Mohammad Rafi} and {Katam Reddy Kiran Moses} and Petersen, Kai and Mantyla, Mika V.},
	month = jun,
	year = {2012},
	pages = {36--42},
}

@inproceedings{daka_survey_2014,
	title = {A {Survey} on {Unit} {Testing} {Practices} and {Problems}},
	isbn = {978-1-4799-6033-0 978-1-4799-6032-3},
	url = {http://ieeexplore.ieee.org/document/6982627/},
	doi = {10.1109/ISSRE.2014.11},
	language = {en},
	urldate = {2022-05-17},
	booktitle = {2014 {IEEE} 25th {International} {Symposium} on {Software} {Reliability} {Engineering}},
	publisher = {IEEE},
	author = {Daka, Ermira and Fraser, Gordon},
	month = nov,
	year = {2014},
	pages = {201--211},
}

@article{storey_who_2020,
	title = {The who, what, how of software engineering research: a socio-technical framework},
	volume = {25},
	issn = {1573-7616},
	shorttitle = {The who, what, how of software engineering research},
	url = {https://doi.org/10.1007/s10664-020-09858-z},
	doi = {10.1007/s10664-020-09858-z},
	language = {en},
	number = {5},
	urldate = {2023-01-30},
	journal = {Empirical Software Engineering},
	author = {Storey, Margaret-Anne and Ernst, Neil A. and Williams, Courtney and Kalliamvakou, Eirini},
	month = sep,
	year = {2020},
	pages = {4097--4129},
}

@article{swillus_sentiment_2023,
	title = {Sentiment overflow in the testing stack: {Analyzing} software testing posts on {Stack} {Overflow}},
	volume = {205},
	issn = {0164-1212},
	shorttitle = {Sentiment overflow in the testing stack},
	url = {https://www.sciencedirect.com/science/article/pii/S0164121223001991},
	doi = {10.1016/j.jss.2023.111804},
	language = {en},
	urldate = {2023-07-31},
	journal = {Journal of Systems and Software},
	author = {Swillus, Mark and Zaidman, Andy},
	month = nov,
	year = {2023},
	pages = {111804},
}

@inproceedings{evans_scared_2021,
	title = {Scared, frustrated and quietly proud: {Testers}’ lived experience of tools and automation},
	isbn = {978-1-4503-8757-6},
	shorttitle = {Scared, frustrated and quietly proud},
	url = {https://dl.acm.org/doi/10.1145/3452853.3452872},
	doi = {10.1145/3452853.3452872},
	language = {en},
	urldate = {2024-01-08},
	booktitle = {European {Conference} on {Cognitive} {Ergonomics} 2021},
	publisher = {ACM},
	author = {Evans, Isabel and Porter, Chris and Micallef, Mark},
	month = apr,
	year = {2021},
	pages = {1--7},
}

@article{rooksby_testing_2009,
	title = {Testing in the {Wild}: {The} {Social} and {Organisational} {Dimensions} of {Real} {World} {Practice}},
	volume = {18},
	issn = {0925-9724, 1573-7551},
	shorttitle = {Testing in the {Wild}},
	url = {http://link.springer.com/10.1007/s10606-009-9098-7},
	doi = {10.1007/s10606-009-9098-7},
	language = {en},
	number = {5-6},
	urldate = {2024-01-16},
	journal = {Computer Supported Cooperative Work (CSCW)},
	author = {Rooksby, John and Rouncefield, Mark and Sommerville, Ian},
	month = dec,
	year = {2009},
	pages = {559--580},
}

@book{hoda_qualitative_2024,
	edition = {1st ed. 2024},
	title = {Qualitative {Research} with {Socio}-{Technical} {Grounded} {Theory}: {A} {Practical} {Guide} to {Qualitative} {Data} {Analysis} and {Theory} {Development} in the {Digital} {World}},
	isbn = {978-3-031-60533-8},
	shorttitle = {Qualitative {Research} with {Socio}-{Technical} {Grounded} {Theory}},
	language = {eng},
	publisher = {Springer International Publishing},
	author = {Hoda, Rashina},
	year = {2024},
	doi = {10.1007/978-3-031-60533-8},
}

@article{kitchenham_principles_2002,
	title = {Principles of survey research: part 3: constructing a survey instrument},
	volume = {27},
	issn = {0163-5948},
	shorttitle = {Principles of survey research},
	url = {https://dl.acm.org/doi/10.1145/511152.511155},
	doi = {10.1145/511152.511155},
	language = {en},
	number = {2},
	urldate = {2025-02-11},
	journal = {ACM SIGSOFT Software Engineering Notes},
	author = {Kitchenham, Barbara A. and Pfleeger, Shari Lawrence},
	month = mar,
	year = {2002},
	pages = {20--24},
}

@inproceedings{straubinger_survey_2023,
	title = {A {Survey} on {What} {Developers} {Think} {About} {Testing}},
	url = {https://doi.org/10.1109/ISSRE59848.2023.00075},
	doi = {10.1109/ISSRE59848.2023.00075},
	booktitle = {34th {IEEE} {International} {Symposium} on {Software} {Reliability} {Engineering}, {ISSRE} 2023, {Florence}, {Italy}, {October} 9-12, 2023},
	publisher = {IEEE},
	author = {Straubinger, Philipp and Fraser, Gordon},
	year = {2023},
	pages = {80--90},
}

@book{linaker_guidelines_2015,
	title = {Guidelines for {Conducting} {Surveys} in {Software} {Engineering}},
	url = {https://portal.research.lu.se/files/6062997/5463412.pdf},
	publisher = {Lund University},
	author = {Linåker, Johan and Sulaman, Sardar and Host, Martin and de Mello, Rafael},
	year = {2015},
}

@article{buth_effectiveness_2023,
	title = {Effectiveness of bicycle helmets and injury prevention: a systematic review of meta-analyses},
	volume = {13},
	issn = {2045-2322},
	shorttitle = {Effectiveness of bicycle helmets and injury prevention},
	url = {https://www.nature.com/articles/s41598-023-35728-x},
	doi = {10.1038/s41598-023-35728-x},
	language = {en},
	number = {1},
	urldate = {2025-02-27},
	journal = {Scientific Reports},
	author = {Büth, Carlson Moses and Barbour, Natalia and Abdel-Aty, Mohamed},
	month = may,
	year = {2023},
	pages = {8540},
}

@incollection{kitchenham_personal_2008,
	title = {Personal {Opinion} {Surveys}},
	isbn = {978-1-84800-043-8 978-1-84800-044-5},
	url = {http://link.springer.com/10.1007/978-1-84800-044-5_3},
	language = {en},
	urldate = {2025-03-04},
	booktitle = {Guide to {Advanced} {Empirical} {Software} {Engineering}},
	publisher = {Springer London},
	author = {Kitchenham, Barbara A. and Pfleeger, Shari L.},
	year = {2008},
	doi = {10.1007/978-1-84800-044-5_3},
	pages = {63--92},
}

@article{dyba_instrument_2000,
	title = {An {Instrument} for {Measuring} the {Key} {Factors} of {Success} in {Software} {Process} {Improvement}},
	volume = {5},
	number = {4},
	journal = {Empir. Softw. Eng.},
	author = {Dybå, Tore},
	year = {2000},
	pages = {357--390},
}

@inproceedings{dyba_factors_2003,
	title = {Factors of software process improvement success in small and large organizations: an empirical study in the scandinavian context},
	url = {https://doi.org/10.1145/940071.940092},
	doi = {10.1145/940071.940092},
	booktitle = {Proceedings of the 11th {ACM} {SIGSOFT} {Symposium} on {Foundations} of {Software} {Engineering} 2003 held jointly with 9th {European} {Software} {Engineering} {Conference}, {ESEC}/{FSE} 2003, {Helsinki}, {Finland}, {September} 1-5, 2003},
	publisher = {ACM},
	author = {Dybå, Tore},
	editor = {Paakki, Jukka and Inverardi, Paola},
	year = {2003},
	pages = {148--157},
}

@article{touretillery_how_2014,
	title = {How to {Measure} {Motivation}: {A} {Guide} for the {Experimental} {Social} {Psychologist}},
	volume = {8},
	issn = {1751-9004, 1751-9004},
	shorttitle = {How to {Measure} {Motivation}},
	url = {https://compass.onlinelibrary.wiley.com/doi/10.1111/spc3.12110},
	doi = {10.1111/spc3.12110},
	language = {en},
	number = {7},
	urldate = {2025-03-20},
	journal = {Social and Personality Psychology Compass},
	author = {Touré‐Tillery, Maferima and Fishbach, Ayelet},
	month = jul,
	year = {2014},
	pages = {328--341},
}

@article{stobie_too_2005,
	title = {Too {Darned} {Big} to {Test}: {Testing} large systems is a daunting task, but there are steps we can take to ease the pain.},
	volume = {3},
	issn = {1542-7730, 1542-7749},
	shorttitle = {Too {Darned} {Big} to {Test}},
	url = {https://dl.acm.org/doi/10.1145/1046931.1046944},
	doi = {10.1145/1046931.1046944},
	language = {en},
	number = {1},
	urldate = {2025-04-09},
	journal = {Queue},
	author = {Stobie, Keith},
	month = feb,
	year = {2005},
	pages = {30--37},
}

@misc{pilkington_us_2024,
	title = {{US} transportation, police and hospital systems stricken by global {CrowdStrike} {IT} outage},
	url = {https://web.archive.org/web/20250306211836/https://www.theguardian.com/technology/article/2024/jul/19/crowdstrike-microsoft-outage},
	urldate = {2025-04-09},
	journal = {The Guardian},
	author = {Pilkington, Ed and Aratani, Lauren},
	month = jul,
	year = {2024},
}

@misc{pariseau_crowdstrike_2024,
	title = {{CrowdStrike} outage underscores software testing dilemmas},
	url = {https://web.archive.org/web/20250226164043/https://www.techtarget.com/searchsoftwarequality/news/366599175/CrowdStrike-outage-underscores-software-testing-dilemmas},
	urldate = {2025-04-09},
	journal = {TechTarget},
	author = {Pariseau, Beth},
	month = jul,
	year = {2024},
}

@misc{icse_msr_2025,
	title = {{MSR} 2025 - {Registered} {Reports} - {MSR} 2025},
	url = {https://web.archive.org/web/20250409135823/https://2025.msrconf.org/track/msr-2025-registered-reports?#Call-for-Registered-Reports},
	urldate = {2025-04-09},
	author = {ICSE},
	month = apr,
	year = {2025},
}

@misc{emse_registered_2025,
	type = {2025-04-09},
	title = {Registered {Reports} {\textbar} {Empirical} {Software} {Engineering} - {An} {International} {Journal}},
	url = {https://web.archive.org/web/20250409135914/https://emsejournal.github.io/registered_reports/},
	urldate = {2025-04-09},
	author = {EMSE},
	month = apr,
	year = {2025},
}

@article{hamilton_elite_2021,
	title = {Elite {Cues} and the {Rapid} {Decline} in {Trust} in {Science} {Agencies} on {COVID}-19},
	volume = {64},
	issn = {0731-1214, 1533-8673},
	url = {https://journals.sagepub.com/doi/10.1177/07311214211022391},
	doi = {10.1177/07311214211022391},
	language = {en},
	number = {5},
	urldate = {2025-04-09},
	journal = {Sociological Perspectives},
	author = {Hamilton, Lawrence C. and Safford, Thomas G.},
	month = oct,
	year = {2021},
	pages = {988--1011},
}

@article{cologna_trust_2025,
	title = {Trust in scientists and their role in society across 68 countries},
	issn = {2397-3374},
	url = {https://doi.org/10.1038/s41562-024-02090-5},
	doi = {10.1038/s41562-024-02090-5},
	journal = {Nature Human Behaviour},
	author = {Cologna, Viktoria and Mede, Niels G.},
	month = jan,
	year = {2025},
}

@article{elali_ai_generated_2023,
	title = {{AI}-generated research paper fabrication and plagiarism in the scientific community},
	volume = {4},
	issn = {26663899},
	url = {https://linkinghub.elsevier.com/retrieve/pii/S2666389923000430},
	doi = {10.1016/j.patter.2023.100706},
	language = {en},
	number = {3},
	urldate = {2025-04-09},
	journal = {Patterns},
	author = {Elali, Faisal R. and Rachid, Leena N.},
	month = mar,
	year = {2023},
	pages = {100706},
}

@article{sharp_models_2009,
	title = {Models of motivation in software engineering},
	volume = {51},
	copyright = {https://www.elsevier.com/tdm/userlicense/1.0/},
	issn = {09505849},
	url = {https://linkinghub.elsevier.com/retrieve/pii/S0950584908000827},
	doi = {10.1016/j.infsof.2008.05.009},
	language = {en},
	number = {1},
	urldate = {2025-04-10},
	journal = {Information and Software Technology},
	author = {Sharp, Helen and Baddoo, Nathan and Beecham, Sarah and Hall, Tracy and Robinson, Hugh},
	month = jan,
	year = {2009},
	pages = {219--233},
}

@article{verner_factors_2014,
	title = {Factors that motivate software engineering teams: {A} four country empirical study},
	volume = {92},
	issn = {01641212},
	shorttitle = {Factors that motivate software engineering teams},
	url = {https://linkinghub.elsevier.com/retrieve/pii/S016412121400020X},
	doi = {10.1016/j.jss.2014.01.008},
	language = {en},
	urldate = {2025-04-10},
	journal = {Journal of Systems and Software},
	author = {Verner, J.M. and Babar, M.A. and Cerpa, N. and Hall, T. and Beecham, S.},
	month = jun,
	year = {2014},
	pages = {115--127},
}

@inproceedings{franca_motivation_2011,
	title = {Motivation in software engineering: a systematic review update},
	isbn = {978-1-84919-509-6},
	shorttitle = {Motivation in software engineering},
	url = {https://digital-library.theiet.org/content/conferences/10.1049/ic.2011.0019},
	doi = {10.1049/ic.2011.0019},
	language = {en},
	urldate = {2025-04-10},
	booktitle = {15th {Annual} {Conference} on {Evaluation} \& {Assessment} in {Software} {Engineering} ({EASE} 2011)},
	publisher = {IET},
	author = {Franca, A.C.C. and Gouveia, T.B. and Santos, P.C.F. and Santana, C.A. and Da Silva, F.Q.B.},
	year = {2011},
	pages = {154--163},
}

@article{misra_identifying_2009,
	title = {Identifying some important success factors in adopting agile software development practices},
	volume = {82},
	copyright = {https://www.elsevier.com/tdm/userlicense/1.0/},
	issn = {01641212},
	url = {https://linkinghub.elsevier.com/retrieve/pii/S016412120900123X},
	doi = {10.1016/j.jss.2009.05.052},
	language = {en},
	number = {11},
	urldate = {2025-04-10},
	journal = {Journal of Systems and Software},
	author = {Misra, Subhas Chandra and Kumar, Vinod and Kumar, Uma},
	month = nov,
	year = {2009},
	pages = {1869--1890},
}

@article{machuca_villegas_perceptions_2022,
	title = {Perceptions of the human and social factors that influence the productivity of software development teams in {Colombia}: {A} statistical analysis},
	volume = {192},
	issn = {01641212},
	shorttitle = {Perceptions of the human and social factors that influence the productivity of software development teams in {Colombia}},
	url = {https://linkinghub.elsevier.com/retrieve/pii/S0164121222001224},
	doi = {10.1016/j.jss.2022.111408},
	language = {en},
	urldate = {2025-04-10},
	journal = {Journal of Systems and Software},
	author = {Machuca-Villegas, Liliana and Gasca-Hurtado, Gloria Piedad and Puente, Solbey Morillo and Tamayo, Luz Marcela Restrepo},
	month = oct,
	year = {2022},
	pages = {111408},
}

@article{machuca_villegas_instrument_2021,
	title = {An {Instrument} for {Measuring} {Perception} about {Social} and {Human} {Factors} that {Influence} {Software} {Development} {Productivity}},
	volume = {27},
	copyright = {https://creativecommons.org/licenses/by-nd/4.0/},
	issn = {0948-6968, 0948-695X},
	url = {https://lib.jucs.org/article/65102/},
	doi = {10.3897/jucs.65102},
	number = {2},
	urldate = {2025-04-10},
	journal = {JUCS - Journal of Universal Computer Science},
	author = {Machuca-Villegas, Liliana and Gasca-Hurtado, Gloria Piedad and Morillo Puente, Solbey and Restrepo Tamayo, Luz Marcela},
	month = feb,
	year = {2021},
	pages = {111--134},
}

@misc{wells_extremeprogrammingorg_2009,
	title = {{ExtremeProgramming}.org},
	url = {https://web.archive.org/web/20250309052557/http://www.extremeprogramming.org/rules/unittests.html},
	urldate = {2025-04-22},
	journal = {Unit tests},
	author = {Wells, Don},
	year = {2009},
}

@misc{swillus_who_2025,
	title = {Who cares about testing?: {Co}-creations of {Socio}-technical {Software} {Testing} {Experiences}},
	shorttitle = {Who cares about testing?},
	url = {http://arxiv.org/abs/2504.07208},
	doi = {10.48550/arXiv.2504.07208},
	urldate = {2025-04-22},
	publisher = {arXiv},
	author = {Swillus, Mark and Hoda, Rashina and Zaidman, Andy},
	month = apr,
	year = {2025},
}

\end{document}